\documentclass[11pt,en,citestyle=authoryear,bibstyle=authoryear]{elegantpaper}

\newcommand{\jelcodes}[1]{\vskip 0.5ex\par
  \noindent\normalfont\bfseries JEL~codes: #1}

\title{Switching between states and the COVID-19 turbulence}
\author{Ilias Aarab}
\institute{European Systemic Risk Board, European Central Bank\footnote{\emph{Disclaimer: This paper should not be reported as representing the views of the European Systemic Risk Board (ESRB) or the European Central Bank (ECB). The views expressed are those of the authors and do not necessarily reflect those of the ESRB or ECB.}}\\University of Antwerp}
\date{November 2020}

\usepackage{amsmath,amssymb}
\usepackage{graphicx}
\usepackage{booktabs}
\usepackage{threeparttable}
\usepackage{subcaption}

\makeatletter
\@ifpackageloaded{hyperref}{
  \hypersetup{colorlinks=true,linkcolor=blue,citecolor=blue,urlcolor=blue}
}{
  \usepackage{hyperref}
  \hypersetup{colorlinks=true,linkcolor=blue,citecolor=blue,urlcolor=blue}
}
\makeatother

\begin{document}

\maketitle

\begin{abstract}
In \textcite{Aarab2020AlignedEconomicIndexStateSwitching}, I examine U.S.\ stock return predictability across economic regimes and document evidence of time-varying expected returns across market states in the long run. The analysis introduces a state-switching specification in which the market state is proxied by the slope of the yield curve, and proposes an Aligned Economic Index built from the popular predictors of \textcite{welch_goyal_2008} (augmented with bond and equity premium measures). The Aligned Economic Index under the state-switching model exhibits statistically and economically meaningful in-sample ($R^2 = 5.9\%$) and out-of-sample ($R^2_{\text{oos}} = 4.12\%$) predictive power across both recessions and expansions, while outperforming a range of widely used predictors. In this work, I examine the added value for professional practitioners by computing the economic gains for a mean-variance investor and find substantial added benefit of using the new index under the state switching model across all market states. The Aligned Economic Index can thus be implemented on a consistent real-time basis. These findings are crucial for both academics and practitioners as expansions are much longer-lived than recessions. Finally, I extend the empirical exercises by incorporating data through September 2020 and document sizable gains from using the Aligned Economic Index, relative to more traditional approaches, during the COVID-19 market turbulence.
\par
\jelcodes{G12, G17, E44, C22, C53}
\end{abstract}

\section{Introduction}
A large literature studies whether aggregate stock returns are predictable (e.g.\ \textcite{fama_french_1988}, \textcite{campbell_shiller_stock_prices_1988}, \textcite{kandel_stambaugh_1996}, \textcite{guo_2002}, \textcite{lewellen_2004}, and \textcite{polk_thompson_vuolteenaho_2006}). However, \textcite{welch_goyal_2008} show that many time-varying forecasting models fail to deliver reliable out-of-sample improvements for market timing based on ex-ante information. Two recurring challenges are model uncertainty (the appropriate specification is unknown ex ante) and parameter instability (estimates vary strongly across sample periods), both of which can materially degrade real-time performance.

In response, more recent work has proposed approaches that explicitly address these issues, including economically motivated restrictions \parencite{pan_pettenuzzo_wang_2018}, forecast combinations \parencite{huang_jiang_tu_zhou_2015}, regime-shift specifications \parencite{hammerschmid_lohre_2018}, and new predictors \parencite{jiang_lee_martin_zhou_2019}. While these contributions often report economically meaningful in-sample and out-of-sample predictability, performance is frequently concentrated in specific episodes, most notably recessions, rather than being robust across the business cycle.

This state dependence is consistent with evidence and theory suggesting that investors process information differently across regimes. For example, \textcite{cujean_hasler_2017} develop an equilibrium model in which investors rely on different forecasting rules and respond to different types of news depending on the prevailing state. \textcite{devpura_kumar_sunila_2018} formally test for time-varying predictability and find that it is both time-varying and predictor-dependent.

Motivated by these findings, \textcite{Aarab2020AlignedEconomicIndexStateSwitching} proposes a simple state-dependent predictive regression in which coefficients are allowed to differ across market states. The state is proxied by the slope of the yield curve: an inversion corresponds to a down state, while a positive slope corresponds to an up state. The model is combined with a single, interpretable predictor, an Aligned Economic Index, constructed from the fundamental predictors of \textcite{welch_goyal_2008} and supplemented by bond- and equity-premium information.\footnote{The 16 fundamental variables are the dividend-price ratio (log), dividend yield (log), earnings-price ratio (log), dividend-payout ratio (log), equity risk premium volatility, book-to-market ratio, net equity expansion, Treasury bill rate, long-term yield, long-term return, term spread, default yield spread, default return spread, inflation, lagged equity premium, and the corporate bond premium. The data can be retrieved from Amit Goyal’s web page at \url{http://www.hec.unil.ch/agoyal/}. A detailed description is provided in \textcite{welch_goyal_2008}.}

The index is constructed using the partial least squares (PLS) approach introduced by \textcite{wold_1975} and refined by \textcite{kelly_pruitt_2013,kelly_pruitt_2015}. \textcite{Aarab2020AlignedEconomicIndexStateSwitching} shows that the resulting Aligned Economic Index ($E^{\text{PLS}}$) outperforms other combination approaches, such as principal components ($E^{\text{PCA}}$) and the forecast-combination approach of \textcite{rapach_strauss_zhou_2010}, as well as a range of widely used predictors.

This study focuses on a practitioner-oriented question: how much economic value does the Aligned Economic Index deliver when used to guide real-time portfolio allocation? To answer this question, I quantify the gains for a mean-variance investor and compare them to standard benchmarks.

\section{Data}
The aggregate stock market return is measured as the excess return: the continuously compounded log return on the S\&P 500 index (including dividends) minus the risk-free rate (proxied by the three-month Treasury bill). By focusing on excess returns, I net out inflation and the level of interest rates, thereby targeting predictability of the real risk premia.

I use the updated dataset of \textcite{welch_goyal_2008} covering 14 widely used fundamental predictors from January 1950 to September 2020, supplemented by bond- and equity-premium measures.

The focus on the post-war sample is motivated by both economic relevance and statistical considerations. First, from an investor’s perspective, it is natural to evaluate predictability in more recent decades: a forecasting model is more compelling if it performs reliably out of sample over the modern period, regardless of its earlier performance \parencite{welch_goyal_2008}. Second, \textcite{lewellen_2004} argues that predictive regressions should be estimated using data after World War II, since pre-1945 return dynamics and predictor properties differ markedly (e.g.\ the extreme volatility of the 1930s affects both the variance and persistence of multiple predictors). I also exclude 1945--1949 because dividend policies around the war era were unusually volatile, which may distort the behavior of dividend-based predictors \parencite{frankfurter_1997}. Finally, reserving the first decade for initial experimentation, the main empirical results use January 1960 to September 2020 (729 months), capturing the initial impact of the COVID-19 turbulence.

All data are monthly, and the analysis focuses on the one-month forecasting horizon. First, the state indicators considered here (e.g.\ business cycle expansions and recessions) typically persist for several months, so longer-horizon regressions may mechanically mix regimes within the forecasting window and blur state-specific behavior. Second, \textcite{cochrane_2011} notes that short-horizon predictability often implies stronger predictability at longer horizons, with related evidence in, for example, \textcite{huang_jiang_tu_zhou_2015} and \textcite{rapach_strauss_zhou_2016}.

To implement regime shifts in real time, I construct an ex-ante state indicator based on the yield-curve slope. This avoids reliance on the NBER recession indicator, which is determined ex post and is therefore not directly usable in a real-time allocation rule.\footnote{The NBER recession indicator can be retrieved from the Federal Reserve Bank of St.\ Louis. \textcite{sander_2018} shows that real-time recession classification can be unreliable, and that misclassifying turning points can lead to substantial losses.} Specifically, the slope is measured as the 10-year Treasury yield minus the 3-month Treasury bill rate (secondary market).

\section{Asset allocation exercise}
To provide a direct measure of the forecasting value of the Aligned Economic Index for economic agents, I evaluate its economic value in a mean-variance portfolio choice problem. The key question is whether the predictive model delivers better investment guidance than a standard baseline forecast.

Consider an investor with a one-month horizon who chooses portfolio weights to maximize expected utility of terminal wealth $W_{t+1}$ conditional on information available at time $t$. Following \textcite{campbell_thompson_2008}, \textcite{neely_rapach_tu_zhou_2014}, \textcite{rapach_strauss_zhou_2016}, and \textcite{sander_2018}, I assume mean-variance preferences in the spirit of \textcite{markowitz_1952}. The investor allocates between the S\&P 500 index and U.S.\ Treasury bills (or a combination of both). The investor’s objective is
\begin{equation}
U_t \;=\; \mathbb{E}_t\!\left(R^{p}_{t+1}\right)\;-\;\frac{\gamma}{2}\,\mathrm{Var}_t\!\left(R^{p}_{t+1}\right),
\label{eq:utility}
\end{equation}
where $R^{p}_{t+1}$ is the simple\footnote{For the asset-allocation analysis, I forecast the simple excess return (rather than the log excess return used in \textcite{Aarab2020AlignedEconomicIndexStateSwitching}).} portfolio return in month $t+1$, and $\gamma$ is the coefficient of relative risk aversion.

I compute the optimal portfolio weights implied by \eqref{eq:utility} using (i) forecasts generated by the Aligned Economic Index under either a one-state regression or a state-switching regression, and (ii) a benchmark model based on the historical mean. I then summarize economic value using differences in certainty-equivalent returns (CERs). As discussed in \textcite{rapach_strauss_zhou_2016}, the CER gain can be interpreted as the annualized portfolio management fee an investor would be willing to pay to access the predictive information relative to the historical-mean benchmark. Appendix~\ref{app:asset_allocation} provides the full derivation and implementation details.

\section{Methodology}
For each predictor-combination method (PLS, PCA, and forecast combination), I compute out-of-sample certainty-equivalent return (CER) gains. Following \textcite{Aarab2020AlignedEconomicIndexStateSwitching}, I produce out-of-sample forecasts in a monthly sample from 1980:01 to 2020:09, beginning after a 20-year training period.\footnote{All index weights are estimated recursively using only information available up to the forecast formation date $t$, so the procedure avoids look-ahead bias.} Each month, I estimate the first two conditional moments of the portfolio return series and compute optimal portfolio weights by maximizing \eqref{eq:utility}. With regard risk preferences, I follow \textcite{huang_jiang_tu_zhou_2015} and set $\gamma=3$.\footnote{\textcite{Aarab2020AlignedEconomicIndexStateSwitching} also reports results for $\gamma=5$ in their work; and show that the qualitative conclusions remain unchanged.} Following \textcite{johnson_2017}, I impose leverage limits (up to 50\%) and rule out short selling, yielding realistic portfolio constraints.\footnote{Short selling by financial institutions has been partially restricted in many countries after the 2008 global financial crisis \parencite{bensoussan_wong_yam_yung_2014}.}

In addition, I restrict month-to-month portfolio adjustment by requiring that the risky-asset position cannot more than double (or be reduced by more than half) from one month to the next, which helps stabilize weights given the well-known sensitivity of mean-variance allocations to return forecasts \parencite{barbara_j_2015}.

I compute CER gains relative to the historical-mean benchmark and annualize them by multiplying by 1{,}200, so they can be interpreted as annual percentage management fees. I also report results net of proportional transaction costs of 50 basis points (bps) per transaction.

As an additional benchmark, I consider a buy-and-hold investor who purchases the market at the start of the out-of-sample period and holds it without rebalancing through the end of the sample.

Finally, I evaluate state dependence by reporting CER gains separately across expansions, recessions, and yield-curve up/down states, mirroring the state-based evaluation in \textcite{Aarab2020AlignedEconomicIndexStateSwitching}. Statistical significance is assessed by testing $H_0: \Delta \mathrm{CER} \le 0$ against $H_D: \Delta \mathrm{CER} > 0$ using the bootstrap procedure of \textcite{mccracken_valente_2018}.

\section{Empirical results}

\subsection{The one-state regression model}
Table~\ref{tab:asset-allocation} depicts the main results. The upper part of Panel A displays results for the one-state regression models, while the lower part depicts results under the state-switching model. The second column presents the difference in CER gains ($\Delta \mathrm{CER}$) without transaction costs, while column five reports the difference in CER gains net of transaction costs. I also report additional portfolio performance measures. Column 3 shows the monthly Sharpe ratio, computed as the mean portfolio return in excess of the risk-free rate divided by the standard deviation of the excess portfolio return. Column 4 reports relative average monthly turnover, where monthly turnover is defined as the percentage of wealth rebalanced at the end of the month. Relative average turnover is computed as average turnover divided by the average turnover of the investor who uses the historical-average forecast.

There are several noteworthy observations. First, the CER for the portfolio based on the historical-average forecast is 7.56\% for January 1980 to September 2020. The $\Delta \mathrm{CER}$ values of all three combination methods are positive under the one-state regression model; however, only $E^{\text{PLS}}$ has a $\Delta \mathrm{CER}$ that is significantly different from zero at the 10\% level (or stronger). In line with these positive certainty-equivalent gains, the Sharpe ratios of the forecasting models all exceed that of the historical average (0.13), with $E^{\text{PLS}}$ and $E^{\text{FC}}$ yielding the highest ratio of 0.16. Average turnover is 2.09\% for the historical-average portfolio. Portfolios based on the forecasting models turn over approximately three to four times more often than the historical-average portfolio. After accounting for transaction costs, these relatively high turnover rates reduce $\Delta \mathrm{CER}$ gains: values turn negative for $E^{\text{PCA}}$ and $E^{\text{FC}}$, and none of the forecasting models produces a statistically significant $\Delta \mathrm{CER}$ different from zero net of costs.

When looking across different states, the historical-average model produces substantial $\Delta \mathrm{CER}$ gains during expansions and up states of 13.00\% and 10.74\%, respectively, but yields negative gains across recessions and down states of -17.46\% and -1.08\%. As expected, the historical-average model is unable to quickly adapt to sharp drops in equity returns during recessions and down states.

The forecasting models exhibit the reversed behavior across market states. All three models fail to generate significant $\Delta \mathrm{CER}$ gains during expansions, with $E^{\text{PCA}}$ and $E^{\text{FC}}$ even failing to outperform the benchmark model. In contrast, during recessions all three models produce statistically significant $\Delta \mathrm{CER}$ values at the 5\% level (or stronger), with $E^{\text{PCA}}$ yielding the highest gains of 5.82\%.

Lastly, note that in great contrast to the three forecasting models, the simple buy-and-hold strategy yields significant $\Delta \mathrm{CER}$ gains of 1.51\% over the full sample period. Moreover, buy-and-hold produces positive $\Delta \mathrm{CER}$ values across expansions (although not significantly), recessions, and up states. During down states, however, buy-and-hold underperforms the historical-average model and produces $\Delta \mathrm{CER}$ gains of -3.90\%.

In conclusion, none of the forecasting models can significantly outperform the naive historical benchmark once transaction costs are imposed, with any positive $\Delta \mathrm{CER}$ gains concentrated around recessions. More surprisingly, among the considered approaches the buy-and-hold strategy yields the highest $\Delta \mathrm{CER}$ value and is the only one to outperform the historical-average forecast over the full sample, but it still fails to do so across all market states. These results echo the earlier findings of \textcite{henkel_martin_nardari_2011}: the short-horizon performance of aggregate return predictors such as the dividend yield and the short rate appears negligible during business-cycle expansions but sizable during contractions.

\subsection{The state switching model}
These results, however, change drastically under the state-switching models (lower part of Table~\ref{tab:asset-allocation}). First, the $\Delta \mathrm{CER}$ values of all three combination methods are significantly positive and economically substantial under the state-switching model, with $E^{\text{PLS}}$ yielding the highest performance of 6.09\%. In line with these high $\Delta \mathrm{CER}$ gains, the state-switching models generate considerably higher monthly Sharpe ratios than the historical-average benchmark (0.13), with $E^{\text{PLS}}$ generating the highest ratio of 0.25, almost double that of the benchmark. After netting transaction costs of 50 bps per transaction, all three models still produce large CER gains in excess of the benchmark model, and all are significantly different from zero at the 5\% level (or stronger). More specifically, $E^{\text{PLS}}$ still yields the highest performance gains net of transaction costs, with $\Delta \mathrm{CER}=5.54\%$, implying that a mean-variance investor would be willing to pay an annual management fee of up to 5.54\% to access the forecasts generated by the aligned, state-switching economic index rather than using the naive historical average.

The improvement from the one-state regression models to the state-switching models is substantial. The average Sharpe ratio increases from 0.15 (the average across the one-state models) to 0.23 under the state-switching model. The average net-of-transaction-costs CER gain increases from -0.03\% under the one-state model to 4.42\% under the state-switching model. This confirms the earlier finding of \textcite{Aarab2020AlignedEconomicIndexStateSwitching} that combining the complementary information in the predictors delivers the best performance when regression coefficients are allowed to vary across states. Most noticeably, the buy-and-hold strategy outperforms the one-state forecasting strategies but fails to do so once state switching is introduced. All three state-switching models generate CER gains far exceeding those of buy-and-hold, with $E^{\text{PLS}}$ producing gains that are almost four times larger than those of buy-and-hold.

The improvement under state switching is also evident across states. First, consistent with the one-state regression results, all three state-switching models significantly outperform the historical average during recessions. Second, while the one-state models fail to outperform the benchmark during expansions, the state-switching models all exhibit positive CER gains, with both $E^{\text{PLS}}$ and $E^{\text{FC}}$ (CER gains of 0.89\% and 1.26\%, respectively) significantly outperforming the historical average at the 5\% level (or stronger).

Across down states, all three models outperform the historical average, with CER gains ranging from 7.09\% ($E^{\text{PCA}}$) up to 9.59\% ($E^{\text{FC}}$). Note how both $E^{\text{PLS}}$ and $E^{\text{FC}}$ under the state-switching model consistently outperform the historical average across all market states. In contrast, even the buy-and-hold strategy cannot significantly outperform the historical average during expansions and generates a negative value of -3.90\% during down states (i.e.\ yield-curve inversions, which most often occur shortly before recessions). These findings mirror the out-of-sample $R^2_{\text{oos}}$ results in \textcite{Aarab2020AlignedEconomicIndexStateSwitching}: under the state-switching model, both $E^{\text{PLS}}$ and $E^{\text{FC}}$ consistently outperform the naive benchmark across market states, with $E^{\text{PLS}}$ delivering the strongest overall performance.

\begin{table}[t]
\centering
\begin{threeparttable}
\caption{Asset allocation results}
\label{tab:asset-allocation}
\scriptsize
\setlength{\tabcolsep}{3pt}
\begin{tabular}{lcccccccc}
\toprule
& \multicolumn{4}{c}{Overall} & \multicolumn{4}{c}{Across states (cost = 50bps)} \\
\cmidrule(lr){2-5}\cmidrule(lr){6-9}
Predictor & $\Delta$CER & SR & Rel.\ avg.\ turnover & $\Delta$CER (50bps) & $\Delta$CER$_{\text{exp}}$ & $\Delta$CER$_{\text{rec}}$ & $\Delta$CER$_{\text{up}}$ & $\Delta$CER$_{\text{down}}$ \\
\midrule
HA         & 7.65 & 0.13 & 2.47 & 7.5   & 13.00  & -17.46 & 10.74 & -1.08 \\
Buy \& hold& 1.35** & 0.16 & 0.00 & 1.51** & 1.10 & 4.09** & 2.39** & -3.90 \\
\addlinespace
\multicolumn{9}{l}{\textit{One-state model}}\\
$E^{\text{PLS}}$ & 1.17* & 0.16 & 5.42 & 0.52 & 0.30 & 5.58*** & 0.93 & -2.00 \\
$E^{\text{PCA}}$ & 0.54  & 0.14 & 8.65 & -0.59 & -1.17 & 5.82*** & 0.38 & 1.46** \\
$E^{\text{FC}}$  & 1.16  & 0.16 & 8.93 & -0.03 & -0.89 & 2.44** & -0.02 & 4.28** \\
\addlinespace
\multicolumn{9}{l}{\textit{State switching model}}\\
$E^{\text{PLS}}$ & 6.09*** & 0.25 & 4.76 & 5.54** & 0.89** & 18.66*** & 2.02** & 9.27*** \\
$E^{\text{PCA}}$ & 4.45**  & 0.22 & 6.82 & 3.58** & 0.20 & 23.81*** & 0.46 & 7.09*** \\
$E^{\text{FC}}$  & 4.71*** & 0.22 & 4.70 & 4.15*** & 1.26** & 2.92*** & 1.50** & 9.59*** \\
\bottomrule
\end{tabular}
\begin{tablenotes}\scriptsize
\item This table reports portfolio performance measures for an investor with mean-variance preferences and relative risk-aversion coefficient of three who monthly allocates between equities and risk-free bills using either an historical average (HA) or predictive regression equity risk premium forecast. The predictors are the aligned economic index constructed with the PLS-method $E^{\text{PLS}}$, the predictor index based on the PCA-method $E^{\text{PCA}}$ and the predictor based on the forecast combination approach $E^{\text{FC}}$ under either the traditional one-state regression model or the state switching model. The $\Delta$CER statistic is the annualized certainty equivalent return gain for an investor who uses the predictive regression forecast instead of the historical average forecast; for the historical average forecast, the table reports the CER level. $\Delta$CER statistics are also reported separately across different states. $\Delta$CER$_{\text{exp}}$, $\Delta$CER$_{\text{rec}}$, $\Delta$CER$_{\text{up}}$, $\Delta$CER$_{\text{down}}$ are respectively the CER return gains across expansions, recessions, up states and down states. Relative average turnover is the average turnover for the portfolio based on the predictive regression forecast divided by the average turnover for the portfolio based on the historical average forecast; for the historical average forecast, the table reports the average turnover level. The ``cost = 50bps'' statistic is the CER gain assuming a proportional transactions cost of 50 basis points per transaction. Statistical significance for $\Delta$CER is based on the p-value of the bootstrap procedure of \parencite{mccracken_valente_2018} for testing the null $H_0: \Delta \mathrm{CER} \le 0$ against the alternative $H_D: \Delta \mathrm{CER} > 0$. ***, ** and * indicate significance at the 1\%, 5\%, and 10\% levels, respectively.
\end{tablenotes}
\end{threeparttable}
\end{table}

\section{Equity weights and cumulative wealth exercise}
\begin{figure}[http]
\centering

\begin{subfigure}[http]{\linewidth}
  \centering
  \includegraphics[width=.8\linewidth]{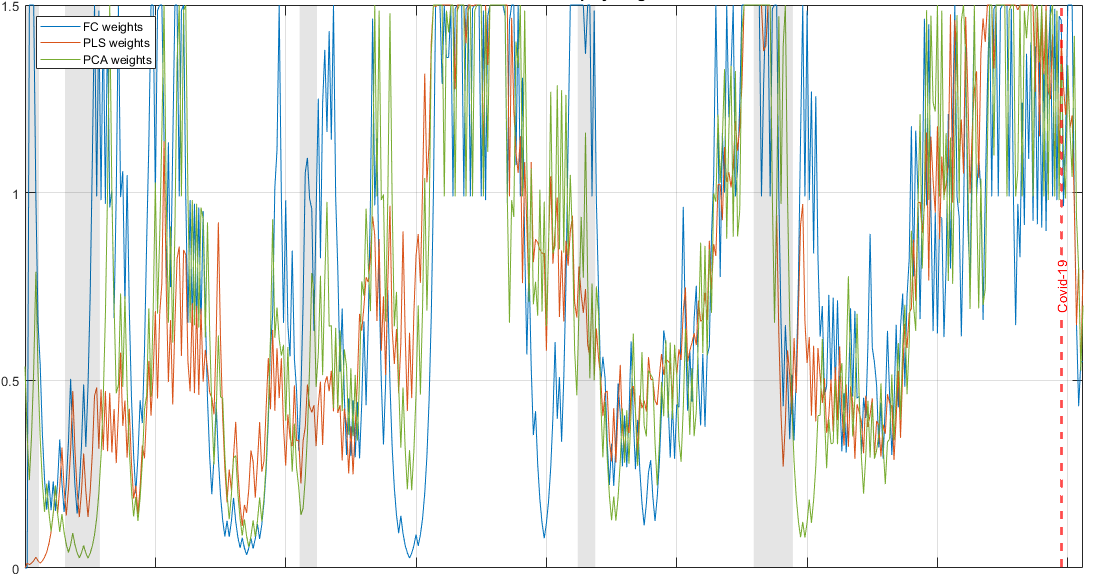}
  \caption{One-state regression model.}
  \label{fig:weights:a}
\end{subfigure}

\vspace{0.6em}

\begin{subfigure}[http]{\linewidth}
  \centering
  \includegraphics[width=.8\linewidth]{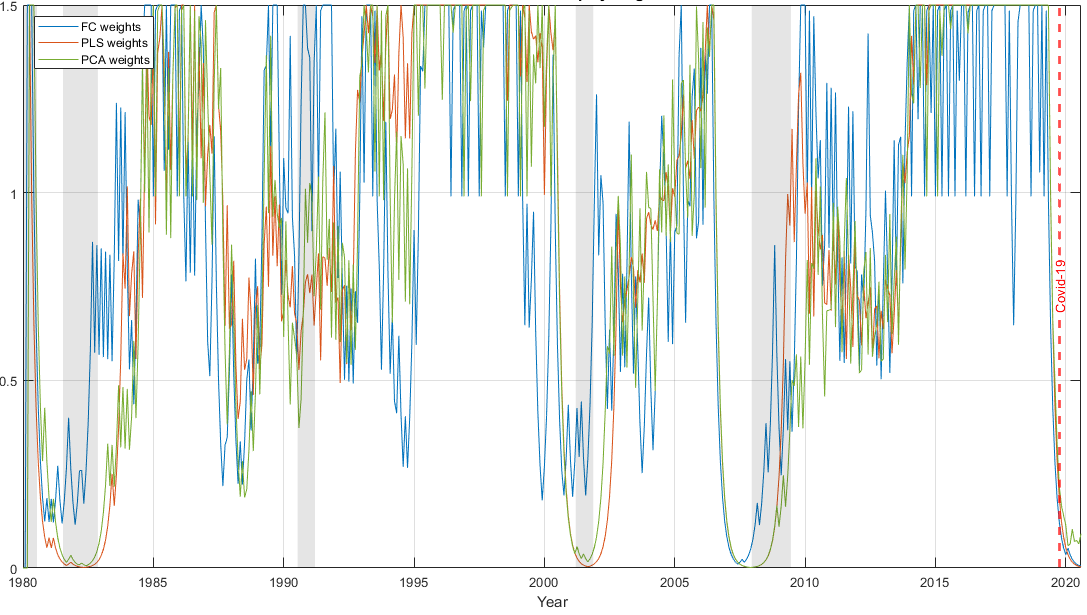}
  \caption{State-switching model.}
  \label{fig:weights:b}
\end{subfigure}

\caption{Asset allocation results, equity weights, January 1980 to September 2020. The upper panel delineates the equity weight for a mean-variance investor with relative risk aversion coefficient of three who optimally allocates across equities and the risk-free asset using a predictive regression excess return forecast based on the one-state regression model of the aligned economic index constructed with the PLS-method $E^{\text{PLS}}$, the predictor index based on the PCA-method $E^{\text{PCA}}$ or the predictor based on the forecast combination approach $E^{\text{FC}}$. The bottom panel corresponds to the same statistics but assumes return forecasts based on the state switching model. The indices, regression coefficients and equity weights are estimated recursively based on information up to the period of forecast formation period $t$ alone. The vertical bars correspond to NBER-dated recessions. The dashed vertical red line indicates the last quarter of 2019.}
\label{fig:weights}
\end{figure}

To better understand the portfolio dynamics implied by the different models, Figure~\ref{fig:weights} plots the equity weights over time. The upper panel shows the one-state models, and the lower panel shows the state-switching models. 

In the one-state models, equity weights are highly volatile, especially for $E^{\text{PCA}}$ and $E^{\text{FC}}$, consistent with the higher turnover reported in Table~\ref{tab:asset-allocation}. More importantly, all three one-state predictors ($E^{\text{PLS}}$, $E^{\text{PCA}}$, and $E^{\text{FC}}$) fail to reduce equity exposure in advance of recessions. For example, at the onset of the subprime crisis all three models are fully invested with equity weights near 150\%. Only after the recession begins do they start reducing exposure, and they do so in a volatile manner, reaching their lowest weights near the end of the recession (roughly 39\% for $E^{\text{PLS}}$, 27\% for $E^{\text{FC}}$, and 16\% for $E^{\text{PCA}}$). In other words, the one-state models adjust too slowly around turning points, remaining heavily exposed during the most adverse period and thereby generating poor CER performance. A similar pattern emerges during the initial COVID-19 market turbulence in early 2020: the models start the episode with substantial equity exposure and then reduce it erratically, ending with average equity weights around 50\% by September 2020, which contributes to sizable portfolio drawdowns. This is consistent with \textcite{sander_2018}, who emphasizes that misclassifying turning points can be costly.

The lower panel of Figure~\ref{fig:weights} shows that state-switching models generate much smoother and more timely adjustments around recessions. The models begin reducing equity exposure before recessions, reach low exposure near the start of recessions, and then gradually increase exposure as conditions improve. For instance, ahead of the subprime crisis the models reduce equity weights from about 150\% at the end of 2006 to near 0\% by the end of 2007, and then rebuild exposure during the recession, reaching local highs near the end of the recession (approximately 130\% for $E^{\text{PLS}}$, 80\% for $E^{\text{PCA}}$, and 70\% for $E^{\text{FC}}$). During the COVID-19 episode, yield-curve inversions in late 2019 and again in early 2020 trigger gradual de-risking, leaving equity weights in the safe single digits by September 2020. As a result, the risky-asset share is materially reduced during the height of the turbulence.

Across expansions, state-switching models can appear more responsive than one-state models and often take larger equity positions. This behavior is consistent with state-dependent slopes: when the model assigns higher expected premia in up states, it rationally tilts more aggressively toward risky assets. By contrast, one-state regressions estimate a single average slope across regimes, which tends to imply more conservative allocations.

As a robustness check and to illustrate the magnitude of the economic gains implied by the state-switching aligned economic index, I also report a cumulative wealth simulation.

Specifically, I consider eight mean-variance investors with identical preferences who each invest an initial wealth of \$1 in a portfolio consisting of T-bills, the market index, or a combination of both. Each investor follows one of the previously discussed strategies (historical-average benchmark, buy-and-hold, $E^{\text{PLS}}/E^{\text{PCA}}/E^{\text{FC}}$ under the one-state model, and $E^{\text{PLS}}/E^{\text{PCA}}/E^{\text{FC}}$ under the state-switching model). The investors allocate from January 1980 through September 2020 using the optimal weights implied by each strategy and reinvest proceeds throughout. All investors have risk aversion coefficient $\gamma=3$.

\begin{figure}[http]
\centering

\begin{subfigure}[t]{\linewidth}
  \centering
  \includegraphics[width=.8\linewidth]{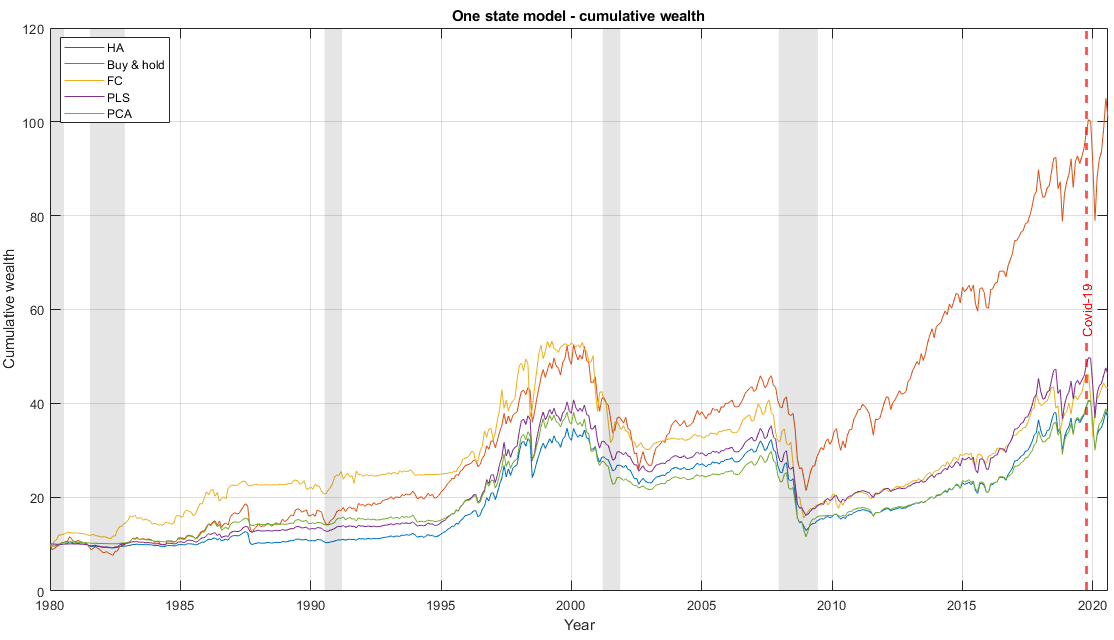}
  \caption{One-state models.}
  \label{fig:wealth:a}
\end{subfigure}

\vspace{0.6em}

\begin{subfigure}[t]{\linewidth}
  \centering
  \includegraphics[width=.8\linewidth]{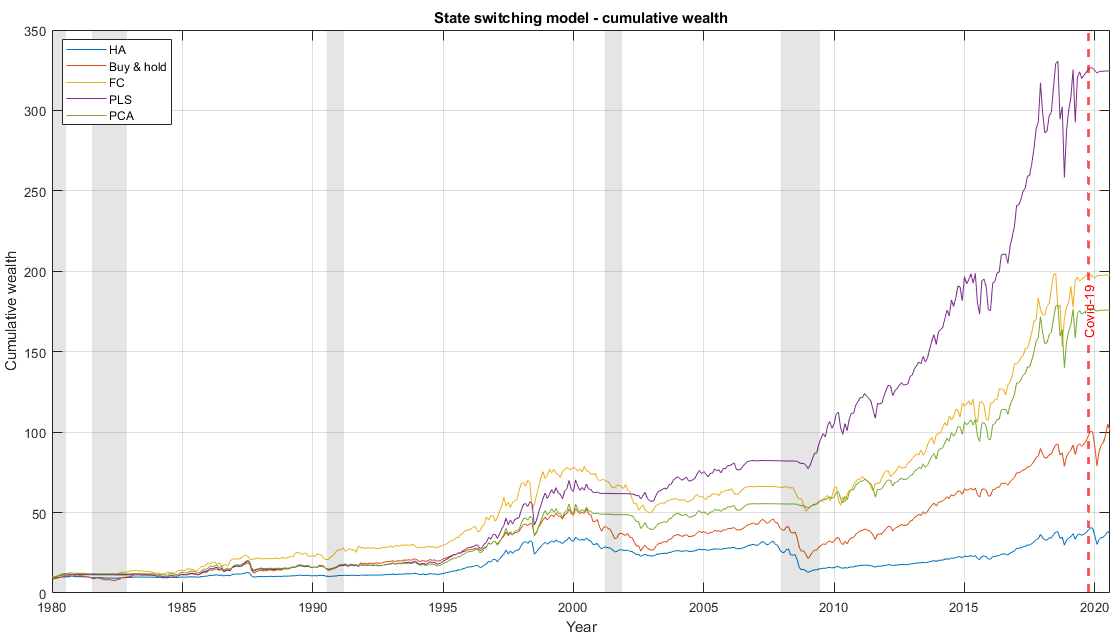}
  \caption{State-switching models.}
  \label{fig:wealth:b}
\end{subfigure}

\caption{Asset allocation results, cumulative wealth, January 1980 to September 2020. The upper panel shows cumulative wealth for a mean-variance investor with relative risk aversion coefficient of three who allocates between equities and the risk-free asset using predictive-regression excess return forecasts based on one-state models: PLS ($E^{\text{PLS}}$), PCA ($E^{\text{PCA}}$), forecast combination ($E^{\text{FC}}$), the historical average (HA), or buy-and-hold. The bottom panel reports cumulative wealth under the corresponding state-switching models. Indices, regression coefficients, and equity weights are estimated recursively using only information available at the forecast-formation period $t$. Vertical bars correspond to NBER-dated recessions. The dashed vertical red line marks the last quarter of 2019.}
\label{fig:wealth}
\end{figure}

Figure~\ref{fig:wealth} plots cumulative wealth. The upper panel shows the one-state models. Consistent with Figure~\ref{fig:weights}, investors experience large drawdowns at the start of recessions because they do not reduce equity exposure quickly enough. The buy-and-hold investor consistently outperforms other one-state strategies during expansions, reaching a terminal wealth of \$102 by September 2020. By contrast, investors using one-state forecasting models end with terminal wealth near \$40, similar to the historical-average investor (\$38). These outcomes mirror Table~\ref{tab:asset-allocation}: among one-state strategies, buy-and-hold performs best and is the only strategy that clearly outperforms the historical-average benchmark.

The lower panel shows cumulative wealth under state switching. All three state-switching strategies are considerably more resilient around recessionary turning points, with far smaller drawdowns. During the COVID-19 turbulence, for example, one-state strategies, the historical-average benchmark, and buy-and-hold all experience pronounced declines, while state-switching strategies remain comparatively stable. Over the full sample, state-switching strategies also accumulate substantially more wealth: terminal wealth reaches \$325 for $E^{\text{PLS}}$, \$197 for $E^{\text{FC}}$, and \$174 for $E^{\text{PCA}}$, far exceeding buy-and-hold (\$102) and the historical-average benchmark (\$38). The outperformance is particularly visible in long expansions. After the 2008 recession, for instance, the $E^{\text{PLS}}$ investor increases wealth from \$91 to \$325 (257\%), compared to \$36 to \$102 (183\%) for buy-and-hold, highlighting the value of the state-switching aligned economic index beyond short recession episodes.

\section{Concluding remarks}
This article evaluates the economic value of a regime-switching return-prediction model that combines state dependence with a new Aligned Economic Index. Consistent with the evidence in \textcite{Aarab2020AlignedEconomicIndexStateSwitching}, I find that using $E^{\text{PLS}}$ within the state-switching framework delivers economically large and statistically significant certainty-equivalent return gains for a mean-variance investor across market states, and outperforms all considered alternatives, including buy-and-hold. These findings matter for practitioners because expansions are long-lived relative to recessions: unlike many predictors whose gains are concentrated in contractions, the state-switching Aligned Economic Index delivers value that is more robust across the business cycle and can be implemented on a consistent real-time basis.

\printbibliography[heading=bibintoc, title=\ebibname]

\appendix
\addappheadtotoc

\section{Asset allocation optimization}
\label{app:asset_allocation}

Consider an economic agent with an investment horizon of one month that wishes to maximize expected utility of terminal wealth $W_{t+1}$ conditional on information up to time $t$. Following \textcite{campbell_thompson_2008}, \textcite{neely_rapach_tu_zhou_2014}, \textcite{rapach_strauss_zhou_2016}, and \textcite{sander_2018}, I assume mean-variance preferences as in \textcite{markowitz_1952}. The investor can allocate between the S\&P 500 index and U.S.\ Treasury bills (or a combination). The utility function is given in \eqref{eq:utility}.

Let $R^{r}_{t+1}$ and $R^{f}_{t+1}$ denote the return on the risky asset (S\&P 500) and the risk-free asset (T-bills) at time $t+1$, respectively. Let $\omega_t$ denote the share invested in the risky asset. The portfolio return is
\begin{equation}
R^{p}_{t+1} \;=\; \omega_t R^{r}_{t+1} + (1-\omega_t) R^{f}_{t+1}.
\label{eq:portfolio}
\end{equation}
Rearranging,
\begin{equation}
R^{p}_{t+1} \;=\; R^{f}_{t+1} + \omega_t R^{e}_{t+1},
\label{eq:portfolio_excess}
\end{equation}
where $R^{e}_{t+1} = R^{r}_{t+1}-R^{f}_{t+1}$ is the excess return.

The investor solves
\begin{equation}
\max_{\omega_t}\ \mathbb{E}_t\!\left(R^{p}_{t+1}\right) - \frac{\gamma}{2}\,\mathrm{Var}_t\!\left(R^{p}_{t+1}\right).
\label{eq:objective}
\end{equation}
Following \textcite{johnson_2017}, I assume that $R^{f}_{t+1}$ is observed at the end of month $t$, so the conditional moments satisfy:
\begin{align}
\mathbb{E}_t(R^{p}_{t+1}) &= R^{f}_{t+1} + \omega_t\,\mathbb{E}_t(R^{e}_{t+1}), \label{eq:mom1}\\
\mathrm{Var}_t(R^{p}_{t+1}) &= \omega_t^2\,\mathrm{Var}_t(R^{e}_{t+1}). \label{eq:mom2}
\end{align}
The first-order condition yields
\begin{equation}
\omega_t^* \;=\; \frac{\mathbb{E}_t(R^{e}_{t+1})}{\gamma\,\mathrm{Var}_t(R^{e}_{t+1})}.
\label{eq:omega_star}
\end{equation}

Because the conditional mean and variance are unknown, they must be estimated. I estimate the conditional mean using the predictive regression forecast from the Aligned Economic Index:
\begin{equation}
\mathbb{E}_t(R^{e}_{t+1}) \;=\; \widehat{r}_{t+1|t},
\label{eq:mean_forecast}
\end{equation}
where $\widehat{r}_{t+1|t}$ is the one-step-ahead forecast produced recursively using only information available at time $t$.

To allow for time-varying volatility, I follow \textcite{campbell_thompson_2008} and estimate $\mathrm{Var}_t(R^{e}_{t+1})$ using a five-year rolling window of historical excess returns:
\begin{equation}
\widehat{\sigma}^{2}_{t+1|t} \;=\; \frac{1}{N-1}\sum_{i=t-N+1}^{t}\left(R^{e}_{i} - \overline{R^{e}}_{t}\right)^2,
\qquad N = 12\times 5 = 60.
\label{eq:rolling_var}
\end{equation}

Thus, the estimated optimal weight is
\begin{equation}
\widehat{\omega}_t^* \;=\; \frac{\widehat{r}_{t+1|t}}{\gamma\,\widehat{\sigma}^{2}_{t+1|t}}.
\label{eq:omega_hat}
\end{equation}
The realized portfolio return is
\begin{equation}
R^{p}_{t+1} \;=\; R^{f}_{t+1} + \widehat{\omega}_t^*\,R^{e}_{t+1}.
\label{eq:portfolio_realized}
\end{equation}

From these returns, the certainty-equivalent return (CER) is
\begin{equation}
\mathrm{CER} \;=\; \mathbb{E}\!\left(R^{p}_{t+1}\right) - \frac{\gamma}{2}\,\mathrm{Var}\!\left(R^{p}_{t+1}\right),
\label{eq:CER}
\end{equation}
which I estimate using unconditional moments of realized portfolio returns. Using unconditional moments is conservative because it requires the estimated conditional moments to be consistent with subsequent return distributions \parencite{johnson_2017}.

For the benchmark, I compute the same allocation rule but replace the predictive mean with the historical mean up to time $t$:
\begin{equation}
\widehat{\omega}^{*,0}_t \;=\; \frac{\overline{r}_t}{\gamma\,\widehat{\sigma}^{2}_{t+1|t}},
\label{eq:omega_hat0}
\end{equation}
where $\overline{r}_t$ is the sample mean up to month $t$.\footnote{Using identical variance estimates ensures that differences in CER gains reflect differences in expected return estimates.}

The benchmark CER is
\begin{equation}
\mathrm{CER}_0 \;=\; \mathbb{E}\!\left(R^{f}_{t+1} + \widehat{\omega}^{*,0}_t R^{e}_{t+1}\right)
- \frac{\gamma}{2}\,\mathrm{Var}\!\left(R^{f}_{t+1} + \widehat{\omega}^{*,0}_t R^{e}_{t+1}\right).
\label{eq:CER0}
\end{equation}

Finally, the CER gain is
\begin{equation}
\Delta \mathrm{CER} \;=\; \mathrm{CER} - \mathrm{CER}_0.
\label{eq:deltaCER}
\end{equation}
As in \textcite{rapach_strauss_zhou_2016}, $\Delta \mathrm{CER}$ can be interpreted as the annualized management fee an investor would be willing to pay for the predictive information embedded in the regression forecast relative to the historical-average forecast.

\end{document}